\documentclass[runningheads]{llncs}
\usepackage{graphicx}
\usepackage{biblatex} 
\usepackage{enumitem}
\setlist[itemize]{label=\textbullet}
\usepackage{diagbox}
\usepackage{makecell}
\usepackage{bbding}
\usepackage{amsmath}
\usepackage{subfigure}
\begin{document}
\title{FRAD: Front-Running Attacks Detection on Ethereum using Ternary Classification Model}
%
%
\author{Yuheng Zhang\inst{1}\and
      Pin Liu \inst{2}\orcidID{0000-0001-5213-2871}\and
      Guojun Wang \inst{1 (}\Envelope\inst{)}\orcidID{0000-0001-9875-4182}\and
      Peiqiang Li\inst{1} \and
      Wanyi Gu\inst{1} \and
      Houji Chen\inst{1}\and
      Xuelei Liu\inst{1}\and
      Jinyao Zhu\inst{1}
}%

%
\institute{\textsuperscript{1}School of Computer Science and Cyber Engineering, Guangzhou University, Guangzhou 510006, China\\
\email{Correspondence to: csgjwang@gzhu.edu.cn}\\
\textsuperscript{2}School of Computer Science and Engineering, Central South University, Changsha 410083, China}

\maketitle              
\begin{abstract}
With the evolution of blockchain technology, the issue of transaction security, particularly on platforms like Ethereum, has become increasingly critical. Front-running attacks, a unique form of security threat, pose significant challenges to the integrity of blockchain transactions. In these attack scenarios, malicious actors monitor other users' transaction activities, then strategically submit their own transactions with higher fees. This ensures their transactions are executed before the monitored transactions are included in the block. The primary objective of this paper is to delve into a comprehensive classification of transactions associated with front-running attacks, which aims to equip developers with specific strategies to counter each type of attack. To achieve this, we introduce a novel detection method named FRAD (Front-Running Attacks Detection on Ethereum using Ternary Classification Model). This method is specifically tailored for transactions within decentralized applications (DApps) on Ethereum, enabling accurate classification of front-running attacks involving transaction displacement, insertion, and suppression. Our experimental validation reveals that the Multilayer Perceptron (MLP) classifier offers the best performance in detecting front-running attacks, achieving an impressive accuracy rate of 84.59\% and F1-score of 84.60\%.

\end{abstract}
\section{Introduction}
In the continuous development of blockchain technology, Ethereum has established its position as a leading public blockchain platform. Its extensive influence has permeated various fields, including but not limited to finance, gaming, and supply chain management \cite{abdulrahman2023ai}. However, the widespread application of this technology has also revealed a series of security challenges, among which the problem of front-running attacks is particularly prominent \cite{daian2020flash}.
Front-running attacks pose a unique security threat to blockchain transactions. In this scenario, malicious hackers engage in the surveillance of transactional activity conducted by other users, then submit their own transactions and pay higher fees to ensure their transactions are executed before the observed transactions are included in the block \cite{piet2022extracting}. This type of attack is particularly common on Ethereum, as the transparency of transactions allows anyone to view unconfirmed transactions in the transaction pool \cite{zhang2023your}.
Moreover, the rise of DeFi (Decentralized Finance) has triggered a surge in Ethereum transaction volume, providing fertile ground for front-running attacks \cite{cernera2023token}. Cybercriminals exploit transaction delays and information asymmetry in the transaction pool to maximize their illicit gains \cite{wang2022impact}, making this a prevalent issue within the Ethereum domain.
The impact of front-running attacks on the Ethereum ecosystem is significant. They increase the transaction risk for ordinary users, potentially leading to a loss of user benefits \cite{xavier2023credible}. Simultaneously, they drive the continuous rise in Ethereum network transaction fees, increasing the cost of transactions for users \cite{bentov2019tesseract}. Ultimately, these attacks may hinder the progress of projects within the Ethereum ecosystem and erode user trust in blockchain technology \cite{zhou2023sok}.

In decentralized applications (DApps) transactions on Ethereum, the problem of front-running attacks is even more severe \cite{xu2023sok}. The susceptibility of DApps to potential threats arises from their inherent openness and transparency, allowing malevolent forces to readily observe and scrutinise transactions. Consequently, these actors can exploit these vulnerabilities to execute assaults through the front-running tactics \cite{eskandari2020sok}. This not only threatens the interests of DApp users but also negatively impacts the stability and sustainability of the entire Ethereum ecosystem. Therefore, researching and addressing the problem of front-running attacks in DApp transactions is of vital importance for protecting user interests, reducing transaction risks, and promoting the healthy development of the Ethereum ecosystem \cite{stucke2022simulation}.
Given the severity of front-running attacks, they have garnered significant attention from industry researchers and developers. A concerted effort is underway to mitigate their impact, with solutions such as batch transactions and Maximum Extractable Value (MEV) already being implemented \cite{weintraub2022flash}. Despite these challenges, the potential of blockchain technology remains vast. The industry continues its pursuit to enhance the security and reliability of this transformative technology.
\begin{figure*}[htbp]
\centerline{\includegraphics[width=1\textwidth]{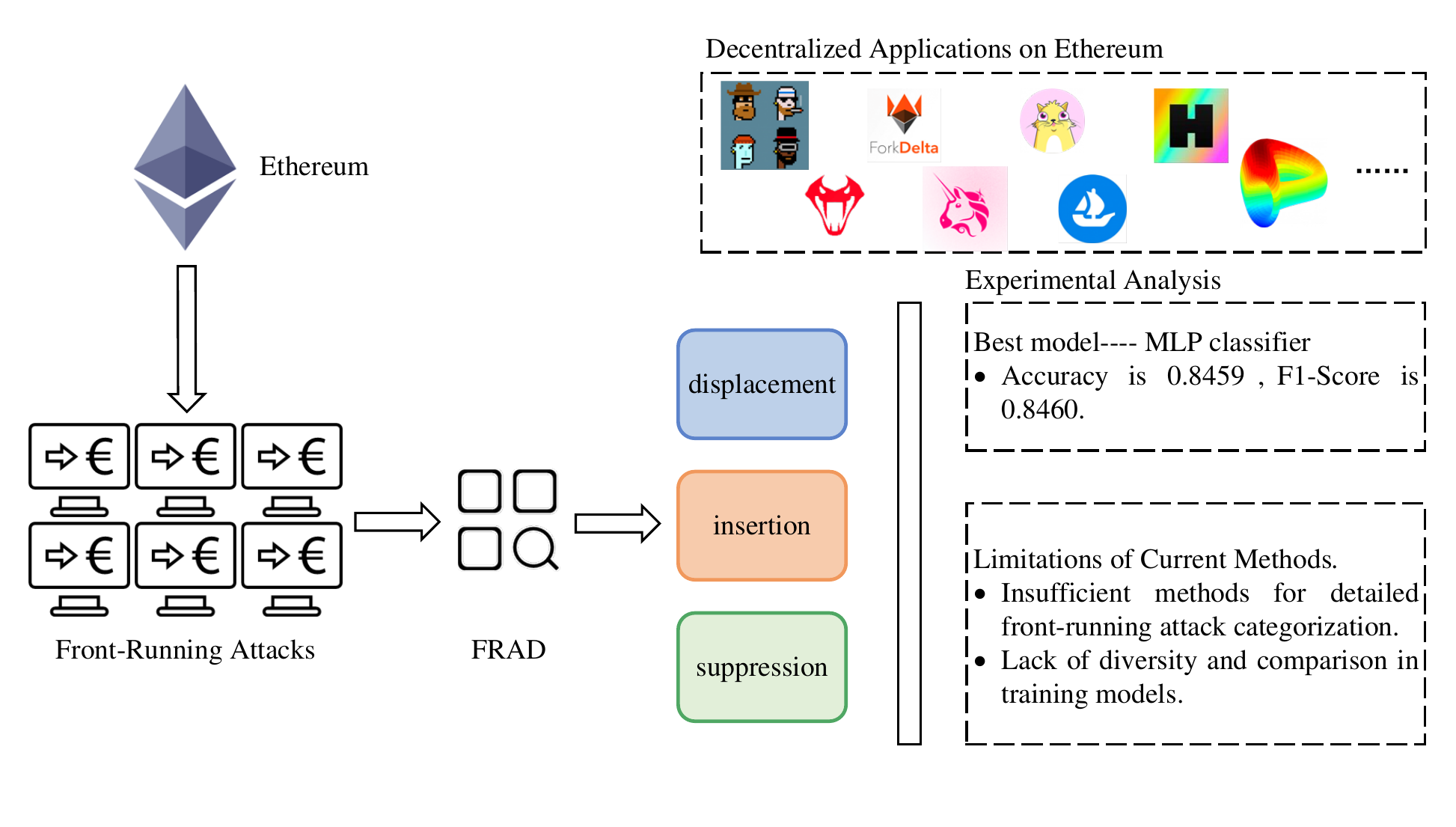}}
\caption{System overview of ternary classification for front-running attacks detection. Initially, the system collects front-running transaction information from decentralized applications (DApps) on Ethereum. Subsequently, the collected transaction information is processed through FRAD to determine whether the transaction belongs to displacement, insertion, or suppression types of attacks.}
\label{fig1}
\end{figure*}

However, the aforementioned issues still remain unresolved, necessitating further exploration and improvement. Front-running attacks can be classified into three basic types: displacement, insertion, and suppression. Each type, while falling under the umbrella of front-running attacks, exhibits distinct behavioral patterns, requiring different countermeasures. Classifying front-running attacks can aid developers in understanding the specifics of the attacks, enabling them to adopt more effective countermeasures. To this end, we built a detection model named FRAD, designed to detect these three categories of front-running attacks, as shown in Fig.~\ref{fig1}.
We utilized data obtained from frontrunner \cite{torres2021frontrunner}. After a series of processing steps, this data was used to train four models: Extreme Gradient Boosting classifier, Gradient Boosting classifier, Random Forest classifier, and Multilayer Perceptron classifier (hereinafter referred to as XGB, GB, RF, and MLP). We then conducted a comprehensive evaluation of the detection results from these four models. The experimental results indicate that among these, the MLP classifier performs the best in detection, achieving an accuracy of 84.59\% and F1-score of 84.60\%.
The primary contributions of this paper are as follows:
\begin{itemize}
\item We propose a detection model named FRAD, designed to detect three types of front-running attacks in the real-world Ethereum network.
\item We utilize Bayesian hyperparameter optimization to enhance the detection performance of our model.
\item We conduct a comprehensive evaluation using four machine learning algorithms on 9798 real-world transactions. Ultimately, we find our FRAD to be effective, with the MLP classifier demonstrating the best detection performance, achieving an accuracy of 84.59\% and F1-score of 84.60\%.
\end{itemize}
     The remainder of this paper is organized as follows:
Section 2 summarizes the related work of this research. Section 3 details the design and implementation of FRAD. Section 4 aims to introduce the evaluation of FRAD. Section 5 provides a summary of the findings from the experiments conducted in Section 4 and outlines future directions.
\section{Related Work}
Maddipati et al. \cite{varun2022mitigating} introduced a scheme for detecting and preventing front-running attacks, which was based on a deep learning model. This model extracted distinct characteristics from each transaction and converted them into a feature vector, which was then used to analyze and determine whether the transaction is a front-running attack. The main focus of \cite{varun2022mitigating} is on utilizing the dataset from \cite{torres2021frontrunner}, in which Christof and others developed a set of tools for measuring and analyzing front-running attacks on Ethereum. The methods outlined above can be utilized for the quantification and evaluation of three distinct forms of front-running attacks. They conducted a large-scale analysis of the Ethereum blockchain and identified 199,725 attacks, resulting in a cumulative profit of over 18.41 million USD for the attackers. It also explores the implications of front-running and reveals that miners benefit from such practices.

Front-running attacks are primarily classified into displacement, insertion, and suppression attacks. In displacement attacks, an attacker monitors a victim's beneficial transactions and publishes their own with higher fees. This gives the attacker priority and profit, while the victim's transaction fails. Struchkov et al. established a displacement front runner model that enhances its priority by initiating similar transactions at higher prices when arbitrage transactions are detected \cite{struchkov2021agent}. However, due to network latency, some arbitrage transactions near the block's end may still succeed, even if they are benign or non-malicious. Insertion attacks involve an attacker monitoring the victim's front-runnable transactions and publishing two transactions with varying fees. The market price fluctuation post front-running transaction completion leads to the victim's transaction price exceeding the pre-attack price. This causes financial loss for the victim, while the attacker profits. To mitigate the adverse effects of these attacks, Patrick et al. presented splitting front-runnable transactions \cite{zust2021analyzing}. Suppression attacks involve an attacker publishing transactions with higher fees to prevent the victim's transactions from being included in the block. These attacks are costly, as the attacker needs to expend significant fees to reach the block's capacity limit \cite{torres2021frontrunner, capponi2022evolution}.

Given the analysis of these three types of front-running attacks, it is imperative to select suitable solutions tailored to each attack type. Therefore, classifying transactions implicated in front-running attacks is of paramount importance. To address this, we will investigate the use of machine learning for enhanced detection and prevention of front-running attacks on Ethereum. Chen et al. proposed a machine learning-based method that uses algorithms such as decision trees, random forests, and gradient boosting \cite{chen2018detecting}. These algorithms predict whether a transaction could be a front-running attack based on transaction features like transaction fees, size, and time. However, most existing methods can only perform binary classification, i.e., categorizing transactions as either normal or front-running attacks. There is a lack of effective methods for further classification of front-running attacks, such as categorizing them as displacement, insertion, or suppression attacks. Despite this limitation, researchers are not solely focused on detecting front-running attacks. In blockchain security, researchers have begun exploring classification techniques to identify and categorize various attacks on Ethereum. For instance, Gu et al. proposed a new method for detecting unknown vulnerabilities on Ethereum using a CNN-BiLSTM model \cite{gu2022detecting}. This model, which belongs to multi-label classification, can detect sequences with unknown vulnerabilities and attacks in Ethereum transaction. Similarly, Li et al. used binary and ternary classification models, combined with a vector weight penalty mechanism to extract operational code features in Ethereum transaction, and then employed three machine learning models to detect unknown attacks and threats \cite{li2022detecting}.

In summary, front-running attacks pose a enormous threat in the blockchain world, and their detection is essential. Current approaches primarily focus on binary classification, distinguishing transactions as either normal or front-running attacks. However, more nuanced classification of front-running attacks, such as categorizing them as displacement, insertion, or suppression attacks, remains an obstacle to overcome. This highlights the significance of further research in this area.
\section{Design and Implementation of FRAD}

\subsection{Framework}
\begin{figure*}[htbp]
\centerline{\includegraphics[width=1.0\textwidth]{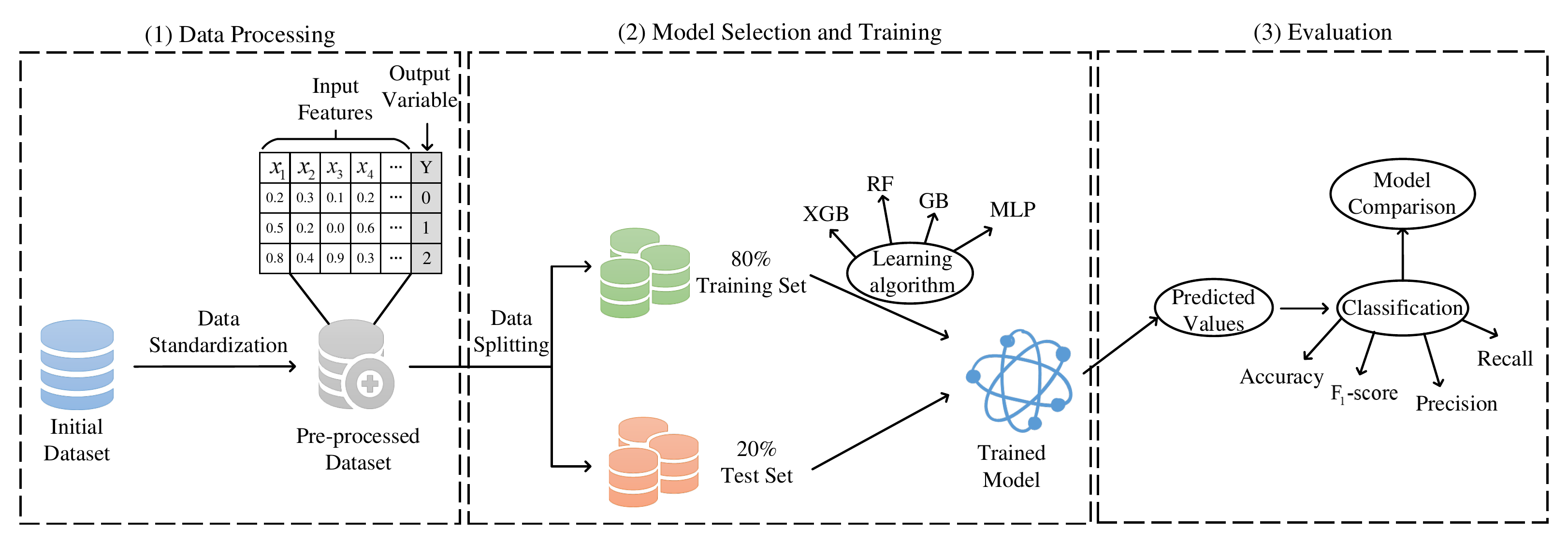}}
\caption{Architecture of FRAD. The learning algorithm box corresponds to XGB, RF, GB, and MLP, which represent Extreme Gradient Boosting Classifier, Gradient Boosting Classifier, Random Forest Classifier, and Multilayer Perceptron Classifier, respectively.}
\label{fig2}
\end{figure*}

As depicted in Fig.~\ref{fig2}, the procedure of FRAD began by standardizing the collected dataset, which was generated using a front-running trading detection technique proposed by Frontrunner Jones \cite{torres2021frontrunner}, to ensure consistent scaling of all features. This particular stage played a critical role in enhancing the suitability of the input for subsequent machine learning algorithms. Following that, we employed Bayesian optimization techniques to refine the model's hyperparameters, aiming to enhance its performance during the training phase. 
After performing data preparation and hyperparameter optimization, we selected four separate machine learning models for the training phase: XGB, GB, RF, and MLP. The models were assigned a classification issue involving three types, which was especially designed to identify three distinct types of front-running trading attacks: displacement, insertion, and suppression. The evaluation of each model's performance was conducted by assessing its detection accuracy, which served as a reliable criterion for facilitating comparisons.

In the final stage of our research, we performed an exhaustive comparison of the four models' performance. Our evaluation was not limited to detecting accuracy; we also considered other performance metrics such as F1-score and precision. This comprehensive evaluation allowed us to understand the strengths and weaknesses of each model thoroughly, thereby enabling us to select the most suitable model for our task.

\begin{figure}[htbp]
\centerline{\includegraphics[width=0.8\textwidth]{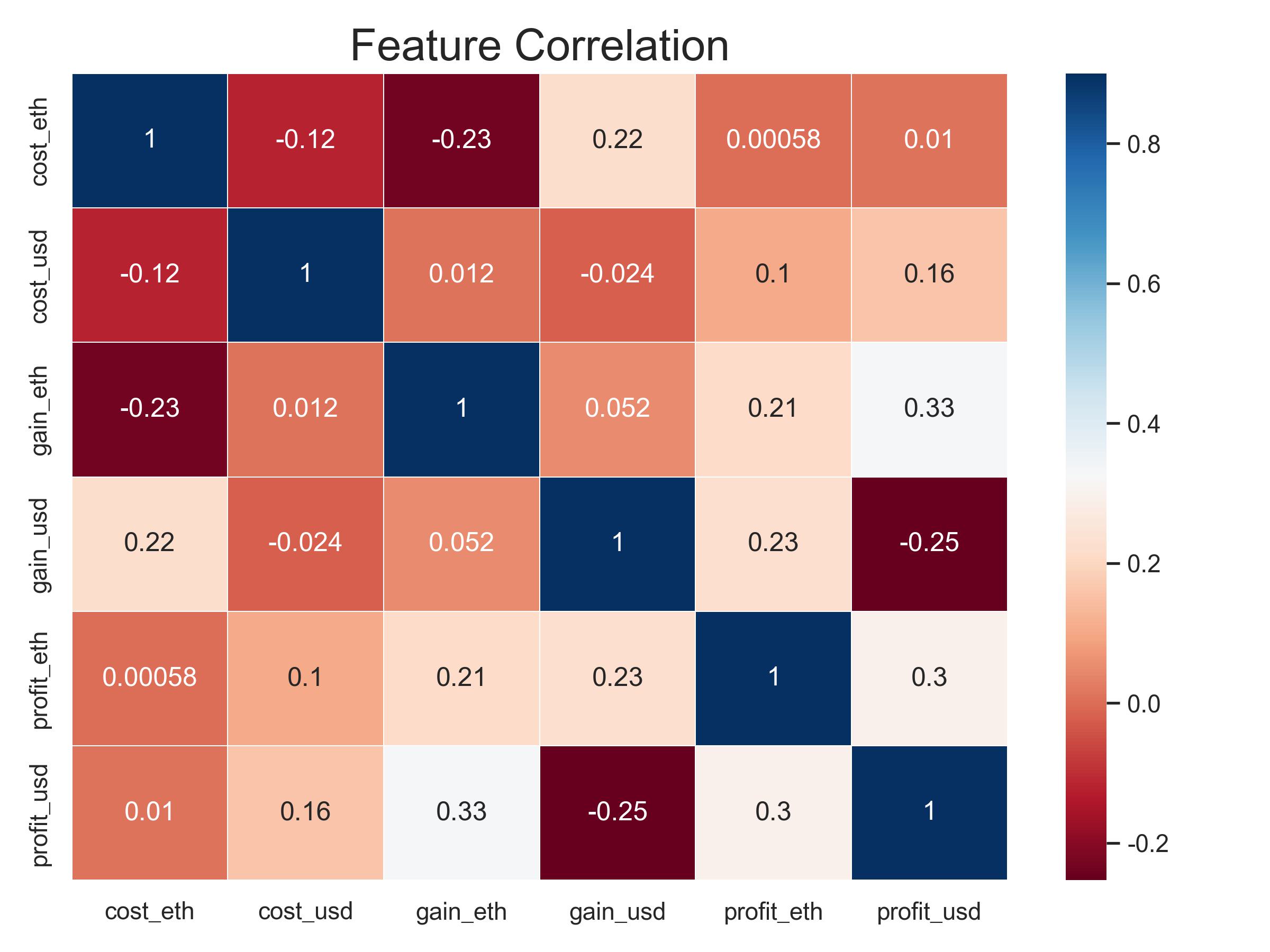}}
\caption{Feature Correlation}
\label{fig3}
\end{figure}
\subsection{Data Processing}
In the data processing phase of our investigation, we imported the dataset and carried out a comprehensive correlation analysis on its features. By creating and visualizing a heatmap, we were able to distinctly depict the interrelations among the various features. The specific outcomes are demonstrated in Fig.~\ref{fig3}. Subsequently, we standardized the dataset associated with front-running trades. This step is pivotal as it harmonizes the scales of diverse features into a uniform range, thereby facilitating superior input for the subsequent machine learning algorithms. This is especially significant for gradient-based optimization algorithms such as linear regression, logistic regression, and neural networks. If the scales of features diverge considerably, it could pose challenges in achieving convergence during the optimization process or lead to convergence to suboptimal solutions. Furthermore, when the data distribution approximates a standard normal distribution across each dimension, data standardization can expedite the learning process, enabling gradient-based optimization algorithms to locate optimal solutions more swiftly.
After the aforementioned processing, as shown in Fig.~\ref{fig2}, the data will be assigned labels of 0, 1, and 2, representing three distinct categories of front-running attack transactions, namely displacement, insertion, and suppression.

\subsection{Model Selection and Training}
After processing the data, we employed Bayesian hyperparameter optimization techniques to pinpoint the best model parameters. Unlike conventional grid search or random search approaches, Bayesian optimization is more efficient in allocating computational resources \cite{wu2019hyperparameter}. It generates a probabilistic model of the objective function and uses this model to select the subsequent parameter for evaluation, thus maximizing computational resource utilization in the pursuit of optimal hyperparameters. We then designated 80\% of the available data as the training set, with the remaining portion assigned as the testing set. Considering the large volume of the front-running trade dataset and the considerable time and computational resources necessary for model training and evaluation, this method is especially suitable for our study. As previously stated, we chose to train four distinct models: XGB, GB, RF, and MLP.

Extreme Gradient Boosting (XGB), a refined variant of the gradient boosting decision tree algorithm, has regularly demonstrated exceptional performance in many machine learning competitions \cite{yu2020copy}. The need to efficiently and effectively detect front-running trade attacks, which entails handling large volumes of transaction data and complex attack patterns, makes XGB an ideal solution due to its competency and versatility. Additionally, the column block parallelization and automatic handling of missing values of the tool further augment its appropriateness for our dataset.

Another model we employed is Gradient boosting (GB), a potent machine learning algorithm that iteratively incorporates new predictive variables to minimize prediction error on the training set \cite{chakrabarty2019flight}. In the context of front-running trading attack detection, the capacity to accurately predict diverse attack patterns is paramount, making GB an apt choice.

Moreover, we leveraged the Random Forest (RF) algorithm, which is an ensemble learning technique based on decision trees. The technique described in the study by Pal improves the precision and consistency of predictions through the construction of numerous decision trees and the subsequent averaging of their outcomes \cite{pal2005random}. Given the need for strong model stability in order to deliver credible forecasts across a variety of trading scenarios, RF's stability and fast training speed make it an excellent candidate.

MLP, commonly known as Multilayer Perceptron, is a supervised learning technique based on feedforward neural networks. Its ability to acquire knowledge and represent intricate nonlinear connections is critical in accurately forecasting diverse sophisticated assault patterns in the detection of front-running attacks \cite{windeatt2006accuracy}.

Consequently, each of these four models is a good option for our experiment since it can handle the challenging three-class problem of front-running attacks detection and has its own distinct capabilities.

\subsection{Evaluation}
During the last phase of our research, an extensive review was conducted to assess the effectiveness of four different models. The evaluation undertaken in this study encompassed more than just assessing the accuracy of detection. It involved a thorough examination of multiple performance indicators, such as the F1-score, precision, and recall. The comprehensive study enabled the attainment of a deep understanding regarding the inherent strengths and weaknesses demonstrated by each model. As a result, the obtained information enhanced our ability to make a judicious selection of the best suitable model for our specific project. Further elaboration on additional information is provided in Section 4.

\section{Evaluation of FRAD}
\subsection{Performance Measure}
In the course of our inquiry, we conducted a comprehensive analysis and evaluation of front-running attacks following the training of four models: XGB, GB, RF, and MLP. Following the completion of the training phase, our attention turned towards an extensive assessment of key performance metrics. These metrics include accuracy, precision, recall, F1-score, as well as various components of the confusion matrix such as true positives (TP), false positives (FP), true negatives (TN), and false negatives (FN). Taking advantage of these indicators is crucial for the comprehensive understanding and assessment of the efficiency of our models in the detection of front-running trading attacks. Accuracy illuminates the model's capacity to accurately categorize all trading behaviours. Precision indicates the percentage of trades predicted as front-running attacks that are truly front-running attacks. Recall demonstrates the proportion of actual front-running attacks that the algorithm correctly detects. The F1-score is a performance evaluation metric that combines precision and recall using a harmonic mean. This metric provides a thorough assessment of performance. The application of a confusion matrix offers a more extensive viewpoint, since it not only emphasizes the accurate categorization of trading behaviours (TP and TN), but also takes into account the instances where misclassification occurs (FP and FN) \cite{visa2011confusion}. This information is of great value in comprehending the effectiveness of our algorithms in detecting front-running trade attacks, as well as for subsequent model optimization. 

Next, a detailed analysis and evaluation of the confusion matrices for the four models will be performed:

\begin{figure}[htbp]
	\centering
	\begin{minipage}{0.49\linewidth}
		\centering
		\includegraphics[width=1\linewidth]{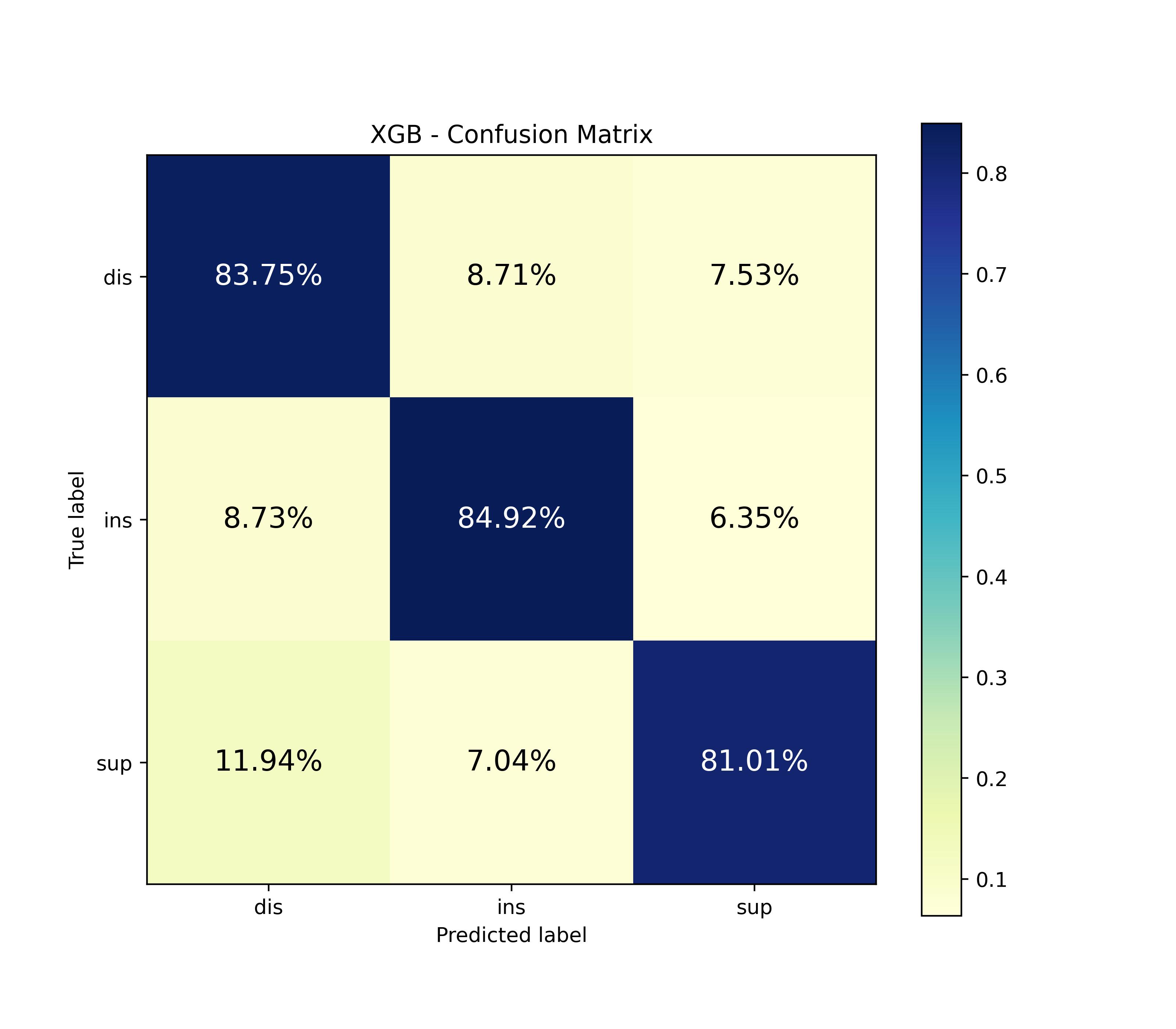}
		\caption{XGB-Comfusion Matrix}
		\label{fig8}
	\end{minipage}
	\begin{minipage}{0.49\linewidth}
		\centering
		\includegraphics[width=1\linewidth]{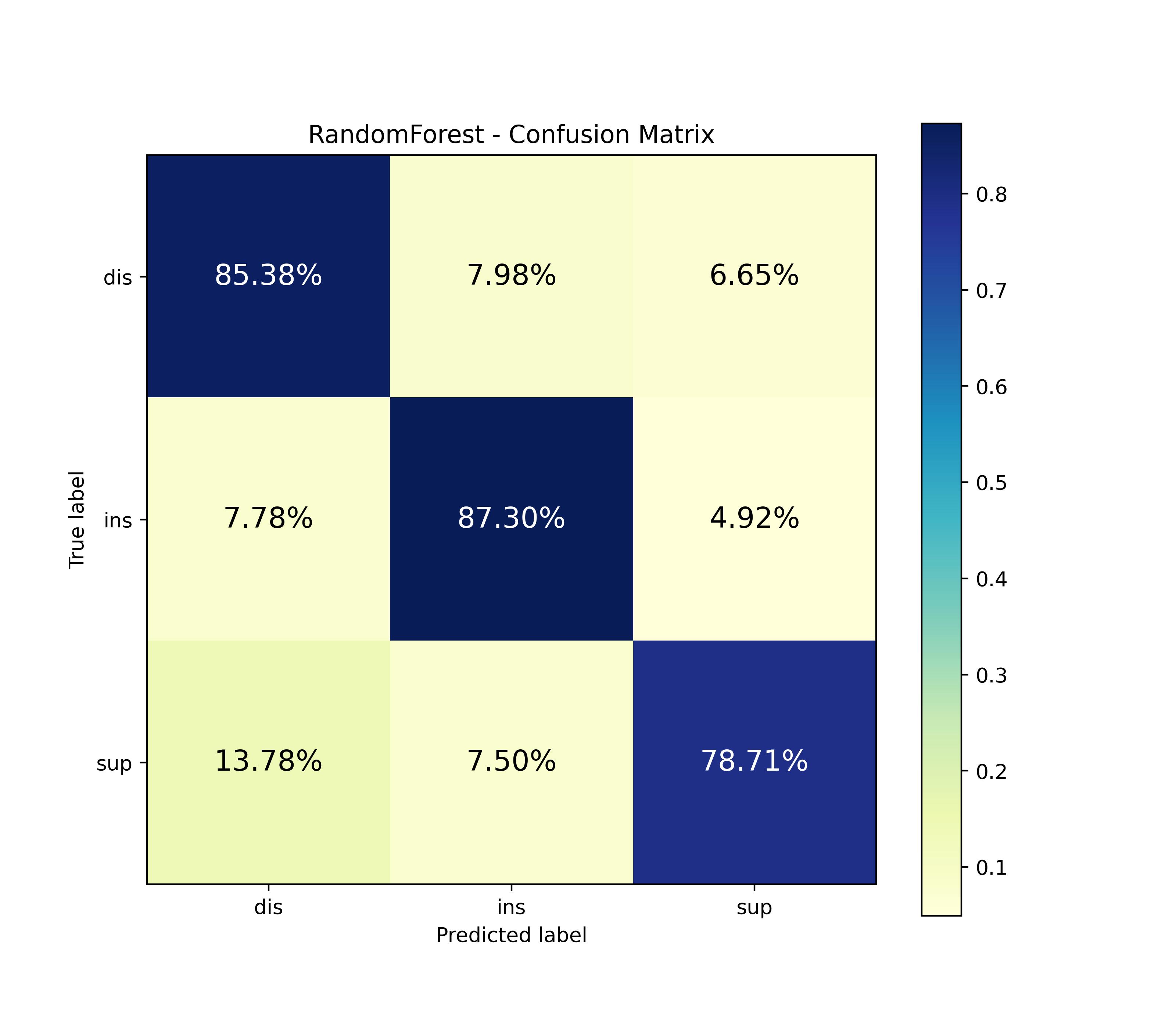}
		\caption{RF-Comfusion Matrix}
		\label{fig9}
	\end{minipage}
	
	\begin{minipage}{0.49\linewidth}
		\centering
		\includegraphics[width=1\linewidth]{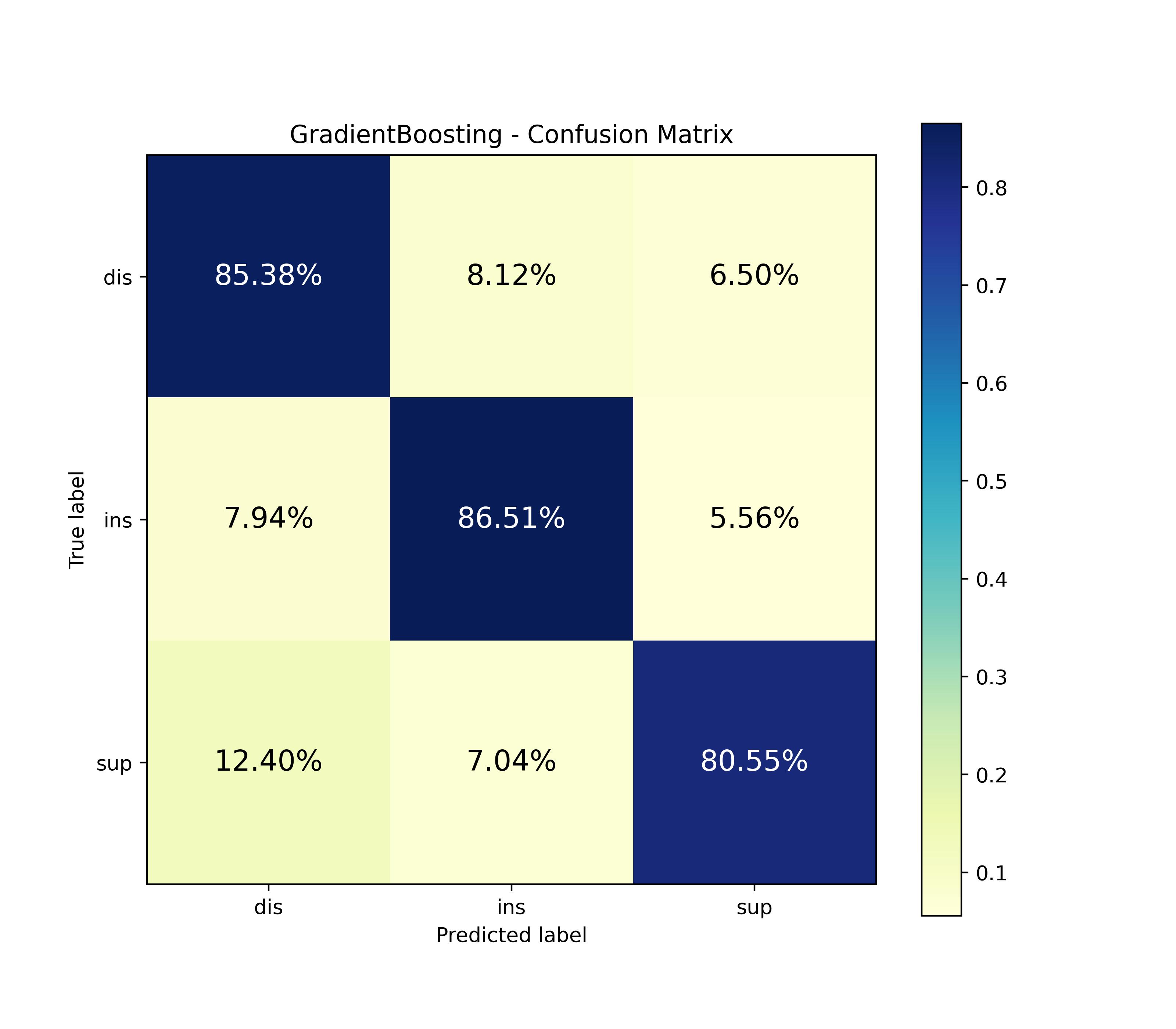}
		\caption{GB-Comfusion Matrix}
		\label{fig10}
	\end{minipage}
	\begin{minipage}{0.49\linewidth}
		\centering
		\includegraphics[width=1\linewidth]{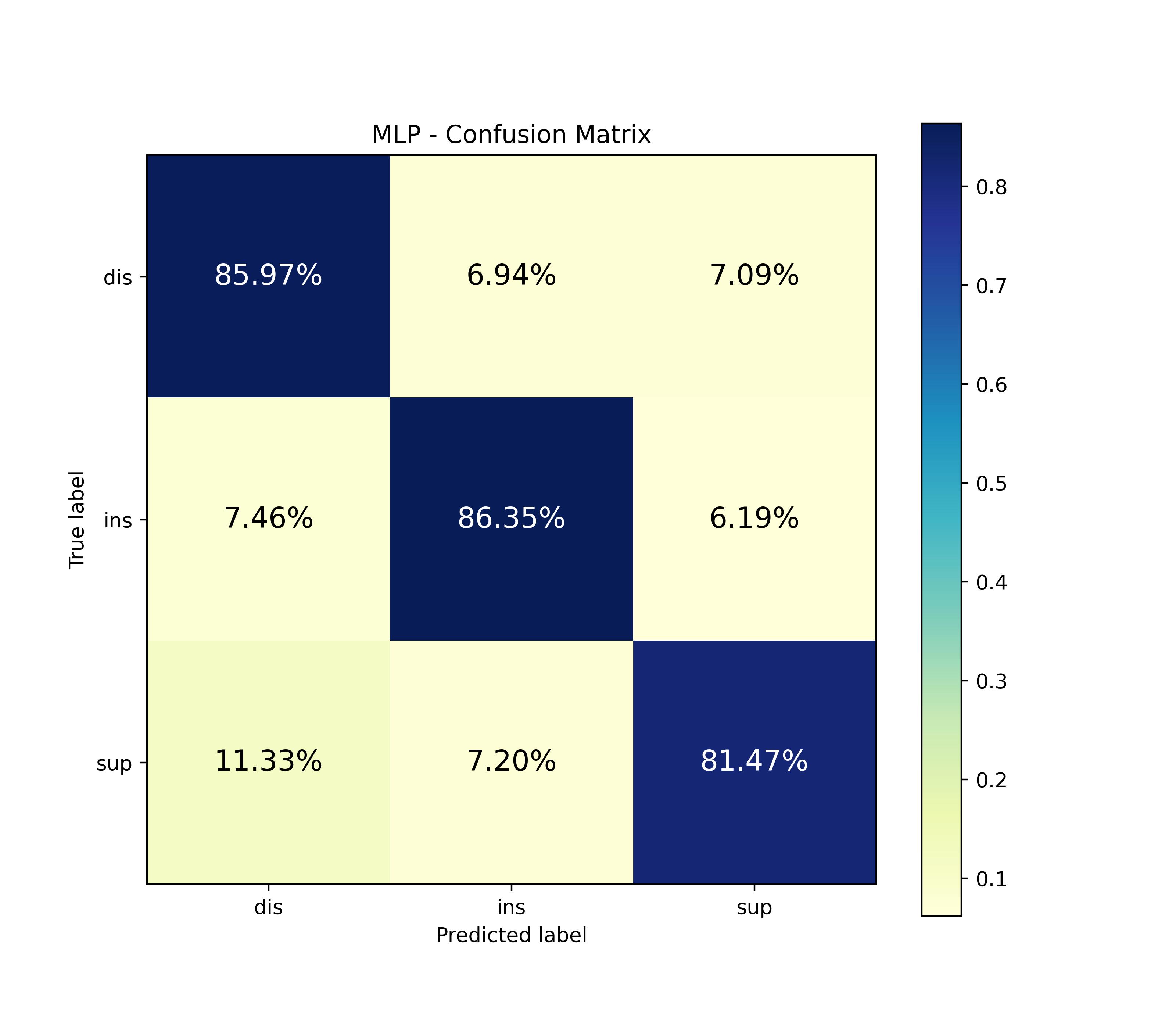}
		\caption{MLP-Comfusion Matrix}
		\label{fig11}
	\end{minipage}
\end{figure}
The XGB model, as illustrated in Fig.~\ref{fig8}, demonstrated a steady performance in managing displacement, insertion, and suppression attacks. The respective accuracy rates for these attacks were 83.75\%, 84.92\%, and 81.01\%. This suggests that the XGB model maintains a consistent level of accuracy in effectively managing these three types of attacks.

On the other hand, the RF model, as shown in Fig.~\ref{fig9}, achieved the highest accuracy rate of 87.30\% when dealing with insertion attacks, the highest among the four models under review. However, despite its impressive performance, the model's capability to identify suppression attacks was somewhat limited, with an accuracy rate of only 78.71\%. Notably, a significant 13.78\% of suppression attacks were mistakenly classified as displacement attacks, which indicates that the RF model's performance may vary when handling different types of attacks.

The GB model's accuracy rates for displacement, insertion, and suppression attacks were 85.38\%, 86.51\%, and 80.55\%, respectively, displayed in Fig.~\ref{fig10}. This means that the GB model demonstrates a consistent high accuracy in acknowledging the three described forms of assaults, albeit with a somewhat lower accuracy in detecting suppression attacks.

In contrast, the MLP model exhibited a very equitable performance, as depicted in Fig.~\ref{fig11}, with accuracy percentages of 85.97\%, 86.35\%, and 81.47\% for the three distinct attack types, respectively. The observed performance of the MLP model suggests its ability to maintain a consistently high level of accuracy across various attack types, without showing disproportionately high or low accuracy rates for any specific attack category.

In conclusion, the analysis of the confusion matrix suggests that the MLP classifier model exhibits a relatively balanced performance across different attack types. Nevertheless, it is necessary to recognize that every model has individual benefits and constraints.  

\subsection{Experimental Results and Analysis}
Following the assessment of the confusion matrix, we proceeded to conduct a comparative analysis of the accuracy, F1-Score, precision, and recall metrics for the four models. These four indicators describe the model's accuracy, balance, precision, and recall rate, which are vital factors when evaluating model performance. By utilizing the metrics mentioned above, a comprehensive evaluation and comparison can be conducted to assess the effectiveness of the four models in detecting front-running attacks. This evaluation will serve as a foundation for choosing the most appropriate model for our research. The calculation formulas employed by them are as follows:
\begin{equation}
	Accuracy = \frac{TP+TN}{TP+TN+FP+FN},
	\label{1} 
\end{equation}
\begin{equation}
	Precision = \frac{TP}{TP+FP},
	\label{2} 
\end{equation}
\begin{equation}
	Recall = \frac{TP}{TP+FN},
	\label{3} 
\end{equation}
\begin{equation}
	F1-score = \frac{2*Precision*Recall}{Precision+Recall}.
	\label{4} 
\end{equation}

\begin{figure}[htbp]
\centerline{\includegraphics[width=0.85\textwidth]{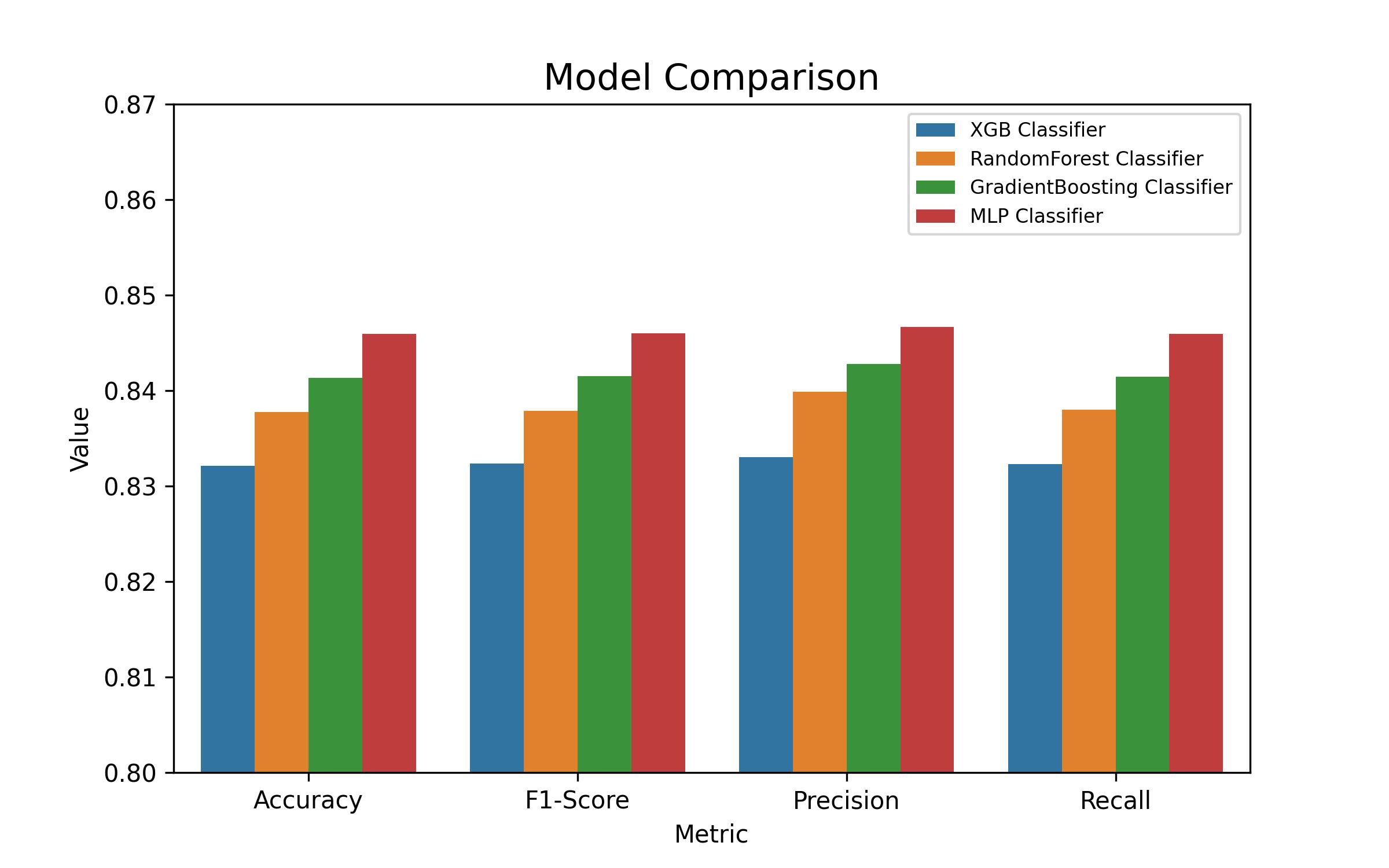}}
\caption{Comprehensive Evaluation of Four Learning Models}
\label{fig12}
\end{figure}

Fig.~\ref{fig12} provided an exhaustive comparative analysis of the four previously mentioned models in terms of their respective performance metrics, namely accuracy, F1-score, precision, and recall. It can be concluded that XGB achieved strong performance in terms of its predictive accuracy, ability to maintain a balance between expected and real positive occurrences, precision in predicting positive outcomes, and the pace at which it accurately predicted positive cases. Nevertheless, RF exhibited slightly better performance when using these measures. GB demonstrated somewhat better performance, with an accuracy of 0.8413, an F1-score of 0.8415, a precision of 0.8427, and a recall of 0.8414. MLP exhibited excellent performance across the four metrics, attaining accuracy, F1-score, precision, and recall values of 0.8459, 0.8460, 0.8466, and 0.8459, respectively. The results of this study indicate that the MLP classifier displays the best performance among the four models.

In summary, all four models performed commendably on the four indicators of accuracy, F1-score, precision, and recall, but the MLP classifier performed best. Specifically, the accuracy of the MLP classifier model reached 0.8459, and the corresponding F1-score achieved 0.8460. The high score obtained demonstrates that the model exhibits exceptional precision and recall when making predictions regarding front-running attacks classifications. In the MLP classifier model, a total of 233 hidden neurons were configured, providing the model with an ample level of complexity to effectively capture subtle patterns present within the data. Concurrently, the starting learning rate for the model was established at 0.0021547501740925594. Because of the model's modest learning rate, it can alter parameters more gently during the learning process, improving the model's learning performance.

\section{Conclusion}
In this work, we present FRAD, a methodology designed exclusively for specific types of attacks occurring within decentralized applications (DApps) on Ethereum, which accurately categorizes front-running attacks into displacement, insertion, and suppression, enabling developers to design proper measures to defend against each type of attack. Additionally, we conducted a comprehensive evaluation of FRAD by employing four machine learning models and assessing their respective metrics. This evaluation serves to showcase the efficacy of FRAD in the detection and analysis of the aforementioned categories of front-running attacks. Our work shows FRAD is anticipated to greatly enhance transaction security within Ethereum's DApps. 

In the future, we have two primary objectives. Firstly, our research aims to utilise a range of open-source technologies to acquire additional transaction data from decentralized applications on the Ethereum platform. This will enhance the comprehensiveness of our dataset and establish a robust empirical foundation for our study. Furthermore, we plan to conduct experiments with ensemble learning approaches, such as Stacking or Voting, to merge the prediction results of the four models employed in this research. Ultimately, we aim to enhance the overall predictive performance.

\section*{Acknowledgments}
This work was supported in part by the National Key
Research and Development Program of China (2020YFB1005804), and in part by the National Natural Science Foundation of China under Grant 62372121.

%
%
%
%
\begin{refcontext}[sorting = none]
\printbibliography

\end{refcontext}

\end{document}